\documentclass[aps,prb,twocolumn,superscriptaddress,10pt]{revtex4-1}
\usepackage[utf8]{inputenc}
\usepackage{libertine}
\usepackage[libertine]{newtxmath}
\usepackage{mathtools,graphicx,microtype,bm}
\usepackage[cal=boondoxo,frak=boondox]{mathalfa}
\usepackage[colorlinks=true,allcolors=blue]{hyperref}

\begin{document}
\title{Charged impurity scattering in two-dimensional topological insulators with Mexican-hat dispersion}

\author{Bagun S.\ Shchamkhalova}
\author{Vladimir A.\ Sablikov}
\email[E-mail:]{s.bagun@gmail.com} 
\affiliation{Kotelnikov Institute of Radio Engineering and Electronics, Fryazino Branch, Russian Academy of Sciences, Fryazino, Moscow District, 141190, Russia}

\begin{abstract}
Scattering by charged impurities is known to mainly determine transport properties of electrons in modern quantum materials, but it remains poorly studied for materials with Mexican hat dispersion. Due to such nontrivial features  as a singular density of states and a ring-shaped Fermi surface, electron-electron interaction and electron transitions between different isoenergetic contours are of key importance in this materials. We show that these factors significantly affect both the spatial profile of the screened potential of Coulomb centers and the dependence of mobility on temperature and electron density. The screened potential is calculated within the random phase approximation. The transport properties are determined without using the usual relaxation time approximation, since the distribution function in energy space is a vector defined by a system of two equations.
\end{abstract}

\maketitle
\section{Introduction}\label{S_1}
Impurity scattering of electrons plays key role in  transport properties of  two–dimensional (2D) materials especially at low temperatures~\cite{Chen2008,babich2025}.
While impurity scattering in materials with parabolic and Dirac dispersions has been studied quite well~\cite{PhysRevX.4.011043}, the charge impurity scattering in the quantum materials with a Mexican hat dispersion (MHD) has not been sufficiently studied yet. Many atomically thin materials such as HgTe/CdHgTe quantum wells~\cite{Krishtopenko}, biased graphene bilayer~\cite{Stauber}, the semiconducting III-VI monochalcogenides, GaS, GaSe, InS, and InSe~\cite{10.1063/1.4928559}, rhombohedral $\alpha$-In$_2$Se$_3$~\cite{kremer2025}, double~\cite{PhysRevB.95.045116, PhysRevLett.119.056803} and triple~\cite{meyer2025} InAs/GaSb quantum wells, transition-metal halogenides~\cite{10.1063/5.0237686}, Sn-doped Bi$_{1.1}$Sb$_{0.9}$Te$_2$S~\cite{PhysRevB.101.121115} with ring-shaped valence or conduction band are investigated recently. The singularity of the density of states (DOS) and intra-contour and inter-contour electron transitions result in unique properties of the polarization and screening of charged impurity potential in materials with a ring-shaped Fermi surface~\cite{PhysRevB.99.085409}. In magnetic field anomalous quantum oscillations of density of states are associated with Landau level quantization in such dispersions~\cite{Alisultanov}.

We are interested in quantum materials with a multi-orbital structure of the band electron states~\cite{PhysRevLett.131.240001} and specifically in the case where the hybridization of orbitals is important. It is in such materials that topological insulators are realized, and their non-trivial quantum-geometric properties attract great interest recently. The multi-orbital structure of the wave functions manifests itself in significant and even dramatic changes of both charge screening~\cite{Sablikov1} and plasma excitations, as studies have shown for some topological materials~\cite{PhysRevLett.119.266804}, graphene~\cite{PhysRevB.106.155422}, and $\alpha-\mathcal{T}_3$ lattice structure~\cite{iurov2021tailoring}.

We study the conductivity of 2D topological insulator with Mexican hat-shaped band dispersion formed due to the inversion of electron and hole bands. This system is interesting for the following reasons. First of all, the basis quantum states have non-trivial quantum-geometric properties, both the quantum metric and the Berry phase, which turn out to be significantly larger under certain conditions~\cite{Sablikov1}. Second, in the energy range between the bottom and the top of the MHD, there are two Fermi contours.  When studying the scattering of electrons in this energy range one needs to consider the scattering within a single as well as between the two Fermi contours. Third, the MHD has two important nontrivial features. The main one, which usually attracts attention, is the Van Hove singularity of the DOS at the MHD bottom. Because of this, the role of electron–electron (e-e) interaction increases significantly and, as a consequence, the conditions for the formation of the ferromagnetic phase~\cite{PhysRevB.75.115425,PhysRevLett.114.236602} and superconducting pairing~\cite{PhysRevLett.56.2732,PhysRevLett.98.167002} are facilitated. Another feature is related to the effective mass of quasiparticles. On the low-wave vector branch of the MHD, the effective mass changes sign with increasing energy from positive near the MHD bottom to negative near the top. This obviously affects the distribution of electron density around the external charge and, consequently, the screened potential, since quasiparticles with negative mass are attracted to the negatively charged center. In particular, due to this feature, quasi-bound states with energy above the MHD top are formed~\cite{SABLIKOV2023115492,SABLIKOV2023129006}. 

The presence of a ring-shaped Fermi surface leads to the appearance of two scattering channels due to inter-contour and intra-contour transitions.
As a result, in addition to the two usually expected Kohn anomalies caused by intra-contour transitions, an additional Kohn anomaly arises, corresponding to the difference in the Fermi vectors of the outer and inner Fermi contours~\cite{Sablikov1}. At low Fermi energies, the inter-contour anomaly is much larger than the other two. The quantum metric significantly affects the amplitude of the singularities. In particular, the inter-contour singularity disappears altogether at high Fermi energies. A three-mode Friedel oscillation structure is formed, the evolution of which with the Fermi energy is determined by the interaction of three main factors: intra-contour and inter-contour transitions, the quantum metric, and the e-e interaction, which plays an important role due to the singularity of the density of states.
 
Another feature of MHD materials is that the momentum in such systems is not a single-valued function of energy and therefore the process of elastic relaxation of the distribution function of non-equilibrium charge carriers takes place through three channels, including inter two isoenergetic Fermi-contour transitions along with intra-contour ones. For this reason, the relaxation time approximation turns out to be unsuitable for describing semiclassical transport. In this paper, we propose a method for solving the Boltzmann equation in this case. The method consists of reducing the Boltzmann integro-differential equation to a system of two linear equations that can be efficiently solved  numerically.

The calculations of the screened impurity potential are performed in the random phase approximation (RPA). The scattering matrix is calculated in the first Born approximation. In the higher approximation, spin-dependent scattering (skew scattering) shows up, and we will consider it separately. All specific calculations are carried out within the frame of the Bernevig-Hughes-Chang (BHZ) model which is rather universal model of topological insulators. Within this model the MHD arises because of the inversion of the electron and hole bands when their hybridization is not too strong. 

In Sec.~\ref{S_potential}, we study the Lindhard polarization function and the screened potential of a point charge and the electron distribution function and conductivity in Sec.~\ref{S_Boltzmann}. In Sec.~\ref{S_conclusion} we discuss the main results and give conclusions.
 
\section{Screened impurity potential}\label{S_potential}

Within the BHZ model~\cite{BHZ} the eigenstates are formed as a result of $sp^3$ hybridization of the electron and hole bands. If the inversion symmetry is not broken, the spin component perpendicular to the layer is a good quantum number, and the total Hamiltonian splits into two Hamiltonians, one for each spin orientation, $s=\uparrow, \downarrow$. The spin-up Hamiltonian is~\cite{BHZ}:
\begin{equation}\label{eq.Hamiltonian}
    H_{\uparrow}= -Dk^2 +
    \begin{pmatrix}
        -M+B \hat{k}^2 & A (\hat{k}_x+i\hat{k}_y)\\
        A (\hat{k}_x-i\hat{k}_y) & M-B \hat{k}^2\\
    \end{pmatrix},
\end{equation}
where the parameter $D$ describe the electron and hole bands asymmetry, $M$, $B$ and $A$ are standard parameters of the model. Their numerical values are known for various materials~\cite{JPSJ.77.031007}.

In what follows we use dimensionless quantities. The values of the energy dimension are normalized to $|M|$, the distance is normalized to $\sqrt{|B/M|}$, the wave vector $k$ is normalized to $\sqrt{|M/B|}$. An important parameter of the model $a=A/\sqrt{|B M|}$ describes the hybridization of the electron and hole bands, $\delta=D/|B|$ describes an electron-hole asymmetry. The MHD is realized when $|a|<\sqrt{2}$. The dispersion relation is 
\begin{equation}\label{spectr}    
\varepsilon_\lambda(k)=-\delta k^2 +\lambda\varepsilon_k,
\end{equation}
where $\varepsilon_k=\sqrt{(1-k^2)^2+a^2k^2}$ and $\lambda=\pm$ is the band index ($\lambda=+$ for conduction band and $\lambda=- 1$ for valence band).
The energy dispersion in conduction band $\varepsilon_+( k)$ is shown in Fig.1 for parameters $a=0.2$ and $\delta=-0.1$. All calculations below are done for these parameters and the Fermi energy is counted from the conduction band minimum $E_0 =0.296$.  

\begin{figure}
    \centerline{\includegraphics[width=0.6\linewidth]{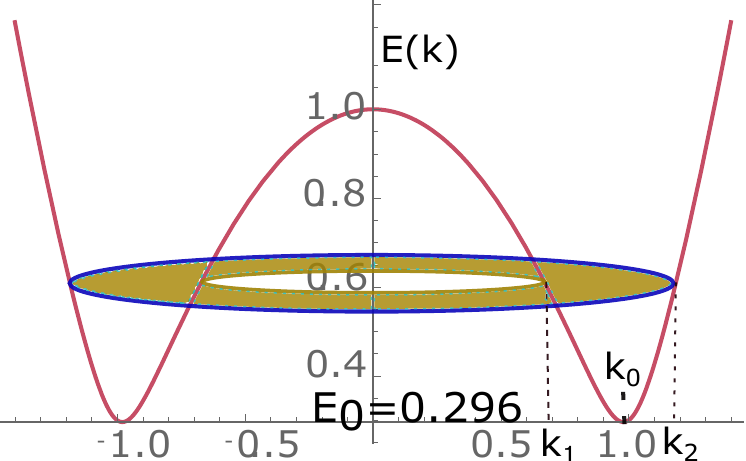}}
    \caption{MHD and ring-shaped Fermi surface for $a = 0.2$ and $\delta = -0.1$, conduction band minimum is at $E_0 =0.296$.}
    \label{fig1}
\end{figure}

The isoenergetic contours in a momentum space are defined by $$k_1(\varepsilon) =\sqrt{(1 - a^2/2 + \varepsilon\delta - \Delta)/(1 - \delta^2)}$$ $$k_2(\varepsilon)=\sqrt{(1 - a^2/2 + \varepsilon\delta + \Delta)/(1 - \delta^2)},$$  where $$\Delta=\sqrt{a^2(a^2/4 - 1) + (\varepsilon + \delta)^2 -\varepsilon a^2\delta}.$$ 
For zero temperature all the momentum states sandwiched between circles of inner radius $k_{F1}=k_1(E_F)$  and the outer radius $k_{F2}= k_2(E_F)$ are occupied. The ring-shaped Fermi surface for the Fermi energy $E_F = 0.3$ is shown in Fig.~\ref{fig1}. 

The spinor $u_{s, \lambda}(\bm{k})$ for spin-up states has the form
\begin{equation}\label{u-sp}
    u_{\uparrow, \lambda}( \bm{k}) = \frac{1}{\sqrt{1+\beta_{\lambda, \bm{k}}^2}}
    \begin{pmatrix}
        1 \\ \beta_{\lambda, \bm{k}} e^{-i\phi_k}  
    \end{pmatrix}\,,
\end{equation}
where 
\begin{equation}
    \beta_{\lambda, \bm{k}}=\frac{a k}{\lambda\varepsilon_k-1+k^2}\,
\end{equation}
and $\phi_k$ is the polar angle of the vector $\bm{k}$.

The wave function of the band state with energy $\varepsilon_\lambda(k)$ is 
\begin{equation}
    |\uparrow\lambda, \bm{k}\rangle =\frac{1}{L}u_{\uparrow\lambda}(\bm{k})e^{i\bm{k r}}\,,
\end{equation}
with the wave vector $\bm{k}$ , $L$ is a normalization length. 

The electron density response to an external potential is calculated in the RPA. In this approximation, the key role is played by the Lindhard polarization function, which describes the density-density response of non-interacting electrons~\cite{giuliani2008quantum}. As a function of the wave vector $\bm{q}$ and the frequency $\omega$, the Lindhard polarization function $\Pi(\bm{q},\omega)$ reads~\cite{giuliani2008quantum,PhysRevLett.119.266804}
\begin{equation}\label{eq.Lidhard_fun}
    \Pi\!(\bm{q},\omega)=2\sum_{\lambda,\lambda'}\int\! \frac{d^2\bm{k}}{(2\pi)^2} \frac{[f(\lambda,\bm{k})-f(\lambda,\bm{k}+\bm{q})] \mathcal{F}_{\lambda,\lambda'}(\bm{k},\bm{k}+\bm{q})}{\hbar \omega +\varepsilon_\lambda(k)-\varepsilon_\lambda(\bm{k}+\bm{q})+i\eta}\,
\end{equation}
 where $f(\lambda,\bm{k})$ is the occupation number, factor 2 is for two spins. 

The form factor $\mathcal{F}_{\lambda,\lambda'}(\bm{k},\bm{k}+\bm{q})$ describes the overlap between the cell periodic parts of the Bloch eigenstates with different quantum numbers:
\begin{equation}\label{eq.overlap_func}
    \mathcal{F}_{\lambda,\lambda'}(\bm{k},\bm{k}+\bm{q})=|u^+_{\lambda}(\bm{k})\, u_{\lambda'}(\bm{k}+\bm{q})|^2\,.
\end{equation}
This function is related to the quantum metric of eigenstates~\cite{resta2011insulating}, $D_{\lambda,\lambda'}(\bm{k},\bm{k'})^2=1-|u^+_{\lambda}(\bm{k}) u_{\lambda'}(\bm{k'})|^2$. An essential feature  is that the dependence of the quantum metric on $k$ and $q$ is of great importance, especially  when $q$ is close to the poles of the integrand in the Lindhard function, Eq.~(\ref{eq.Lidhard_fun}), when the overlap function can significantly change the value of the integral and even affect the presence of the Kohn anomaly.

The presence of two Fermi contours, the space between which is filled with electrons, gives rise to three singularities of the Lindhard polarization function. In addition to the two Kohn anomalies at $q_1=2k_{F1}$ and $q_2=2k_{F2}$, arising from electron transitions within each of the two Fermi contours, there is another singularity at the wave vector $q_0=k_{F2}-k_{F1}$, corresponding to transitions between the Fermi contours. This last singularity turns out to be much larger than the other two in amplitude when the Fermi energy is near the bottom of the MHD. This is obviously due to the divergence of the density of states at the bottom of the band.

The overlap function representing the quantum metric strongly affects the Kohn singularities. Due to reduction of the overlap function, the $q_0$-singularity, which is very large when the Fermi energy is close to the MHD bottom, rapidly decreases in magnitude and disappears  as the Fermi level increases. Quantum geometry also affects the two other singularities at $q_1$ and $q_2$. They are significantly reduced in magnitude, especially the $q_2$-singularity, which practically disappears when $\varepsilon_F$ is close to the MHD bottom. Thus the structure of the polarization function is determined not only by the dispersion of band electrons but also by the quantum metric of the band states~\cite{Sablikov1}.

Within the random phase approximation, the 2D Fourier transform of the screened impurity Coulomb potential equals: 
$$V_{RPA}(q)= \frac{\tilde{V}_q}{1-C_q\Pi(q,0)/q}=\frac{ZC_q}{q-C_q \Pi(q,0)},$$
where $\tilde{V}_q = ZC_q/q$ is 2D Fourier transform of the bare impurity Coulomb potential $V(r)=e^2Z/ \epsilon_0r$,  the impurity is supposed to be a point charge $Z$ at position $\bm{r} = 0$, and $C_q=2\pi e^2/\epsilon_0\sqrt{|MB|}$. Below we omit the index RPA and set $V_q= V_{RPA}(q)$.

Due to singularities of the polarization function the screening is very peculiar. The 2D Fourier transform of the screened impurity Coulomb potential strongly depends on the wave vector and changes significantly depending on the carrier density and temperature.

\begin{figure}\label{fig2}
    \centerline{\includegraphics[width=0.8\linewidth]{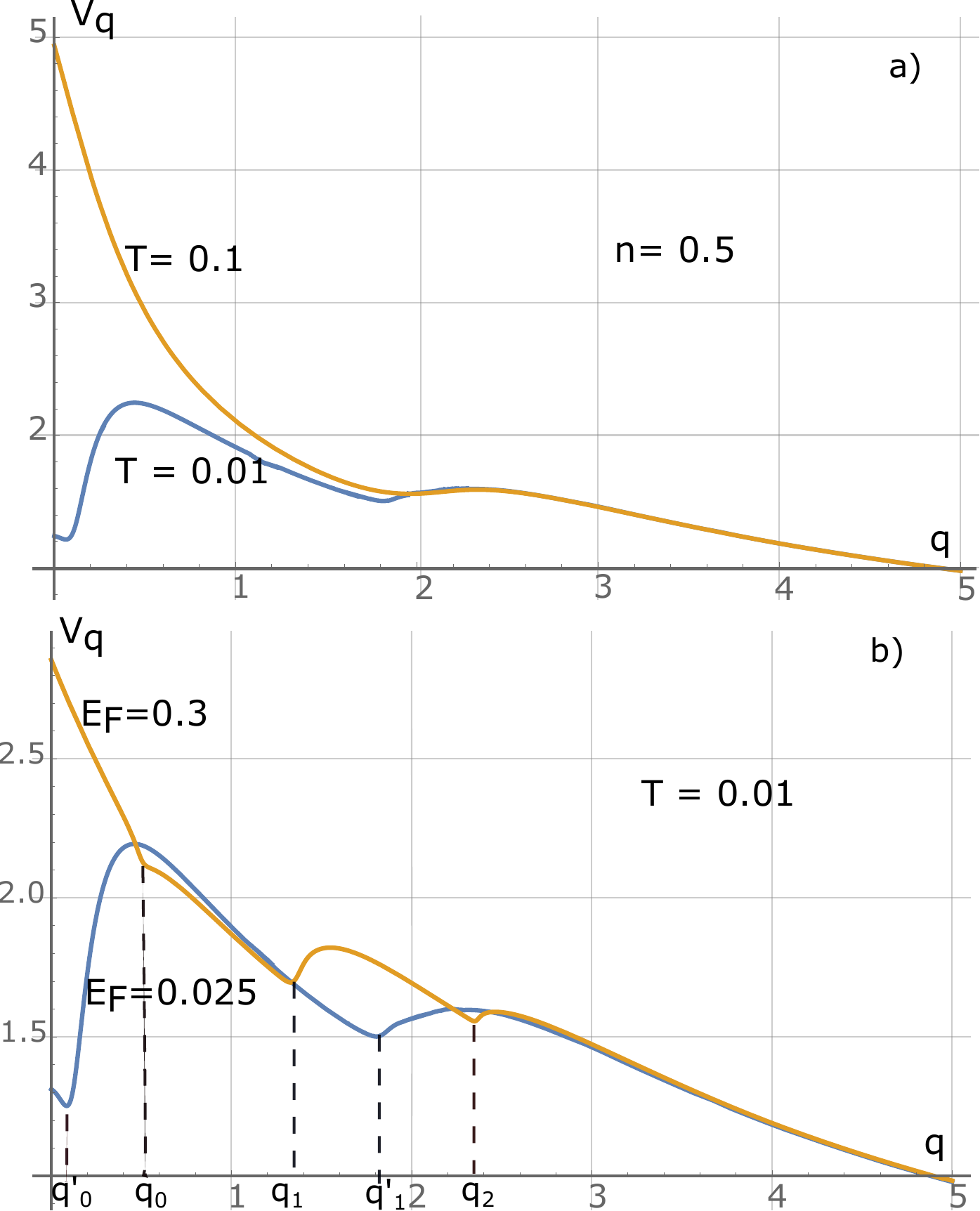}}
    \caption{Wave vector dependence of the screened charge impurity potential: a) at the electron density $n =0.5$ for two values of temperature and 
		b) for two Fermi energies at the same temperature $T = 0.01$. }
 \end{figure}

Fig.2a shows the dependence of $V_q$ on the wave vector at an electron density of $n = 0.5$ for two different temperatures indicated next to the curves. Fig. 2b presents the same dependence for two Fermi energies at a fixed temperature  $T = 0.01$. For the Fermi energy $E_F=0.3$ two anomalies corresponding to wave vectors $q_1 = 2k_{1F}$ (the intra-contour transitions for inner Fermi contour with $k_{1F} = 0.68$)  and $q_2 = 2k_{2F}$ (the intra-contour transitions for the outer Fermi contour with  $k_{2F} = 1.18$) are strong, the anomaly corresponding to their difference $q_0 = k_{2F}-k_{1F} = 0.5$ is weaker. For lower Fermi energy $E_F=0.025$ the wave vectors of the inner ($k'_{1F} = 0.925$) and outer Fermi contours ($k'_{2F}=1.03$) are close, their difference being $q'_0 = k'_{2F}-k'_{1F} = 0.085$. When the Fermi energy is near the bottom of the MHD, the singularities at $q'_1$ and $q'_2$ are significantly reduced in magnitude as an affect of a quantum geometry, the $q'_2$-singularity practically disappears. At the same time another peculiarity at the wave vector $q'_0= 0.085$, corresponding to transitions between the Fermi contours is much larger due to the divergence of the DOS at the bottom of the band.  

Our results demonstrate that reducing the temperature or electron density enhances the screening of the small-angle scattering potential, making the features near the difference of the inner and outer Fermi wave vectors more pronounced.

Scattering matrix elements are:
\begin{equation}
\begin{split}
    \langle\lambda', \bm{k}'|V(r)|\lambda, \bm{k}\rangle =\frac{1+\beta_{\lambda', \bm{k}'}\beta_{\lambda, \bm{k}}e^{i(\phi_k'-\phi_k)}}{\sqrt{(1+\beta_{\lambda', \bm{k}'}^2)(1+\beta_{\lambda, \bm{k}}^2)}}\\\times\int\frac{d^2r}{(2\pi)^2}V(r)e^{i(\bm{k}'-\bm{k})\bm{r}}\equiv\frac{1+\beta_{\lambda', \bm{k}'}\beta_{\lambda, \bm{k}}e^{i(\phi_k'-\phi_k)}}{\sqrt{(1+\beta_{\lambda', \bm{k}'}^2)(1+\beta_{\lambda, \bm{k}}^2)}}V_{\bm{q}},
\end{split}		
\end{equation}
where $\bm{q}=|\bm{k}'-\bm{k}|=\sqrt{k^2+k'^2-2kk'\cos(\phi_k'-\phi_k)}$. 

Spin indexes are omitted here and below because only spin-independent processes for spin up electrons are considered.
\begin{figure}\label{fig3}
    \centerline{\includegraphics[width=1.\linewidth]{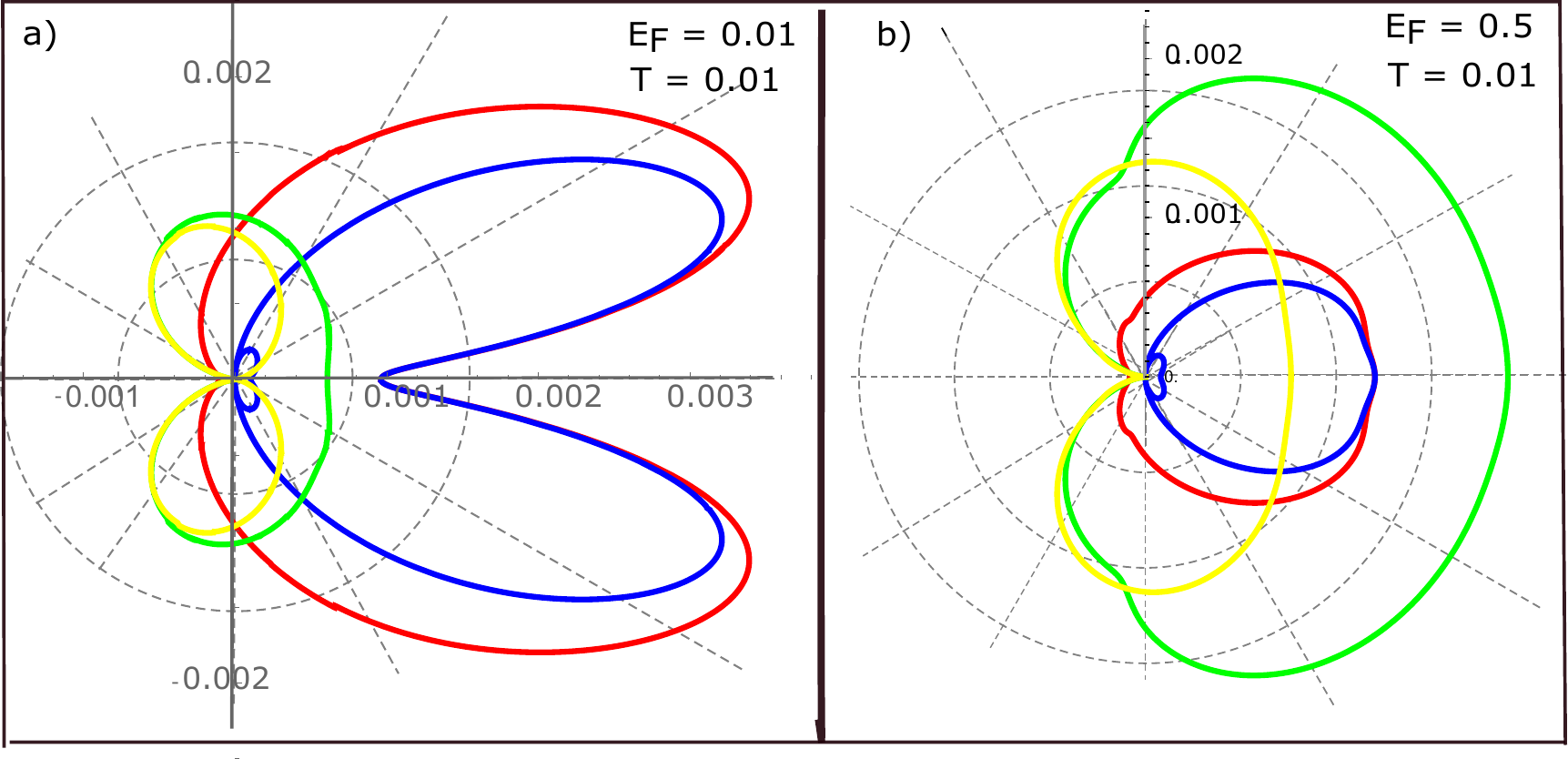}}
    \caption{Angular dependencies of intra-contour (green and yellow) and inter-contour (red and blue) scattering probabilities for $E_F=0.01$ (a) and $E_F = 0.5$ (b).}    
\end{figure}
The matrix elements of intra- and interbranch scattering in MHD systems exhibit a strong dependence on the scattering angle, the dependence changing significantly with the Fermi energy and temperature.
This occurs not only because of the angular dependence of the spinors and the change in the transferred wave vector with the scattering angle, but also because of the very peculiar dependence of the Fourier components of the screened potential on the wave vector, which changes with the carrier density and temperature. 

Fig.~3 shows the squared absolute values of the scattering matrix elements as a function of the scattering angle for two different Fermi levels. The red  and blue curves show $|\langle \bm{k}_1|V| \bm{k}_2\rangle |^2$ and $|\langle \bm{k}_2|V| \bm{k}_1\rangle |^2\cos\phi$ correspondingly (inter Fermi contours for a given Fermi energy), the yellow curves show  $|\langle \bm{k}_1|V| \bm{k}_1\rangle |^2(\cos\phi-1)$ for the intra  first contour scattering, and the green one the same for the second contour. It is seen that the contribution of intracontour and intercontour scattering depends on the Fermi energy: at low energies, intercontour scattering is more intense, while at higher energies, intracontour scattering becomes larger.

\section{Boltzmann equation and current}\label{S_Boltzmann}

We solve the Boltzmann equation for homogenous electron gas:
\begin{equation}\label{Boltz}
\dot{\bm{k}} \frac{\partial f(\lambda,\bm{k})}{\partial\bm{k}}=St[f].
\end{equation}
Here $\dot{\bm{k}} =-(e/\hbar)\bm{E}$ and the scattering term
\begin{equation}
\begin{split}
St[f]=\frac{2\pi}{\hbar}\sum_{\lambda'=\pm}\int\frac{d^2k'}{(2\pi)^2}|\langle\lambda',\bm{k}'|V|\lambda,\bm{k}\rangle |^2[f(\lambda',\bm{k'})-f(\lambda,\bm{k})]\\ \times\delta[\varepsilon_\lambda(k)-\varepsilon_\lambda'(k')].
\end{split}
\end{equation}
The scattering integral is simplified by dividing the k-space into regions $k<k_0$ and $k>k_0$, where $k_0$ is the wave vector of the MHD minimum $E_0$. Thus, the momentum is a single-valued function of energy in each region. Explicitly  keeping track of the two branches of the dispersion,
\begin{equation}
\begin{split}
\int d^2k' g(k')\delta[\varepsilon_\lambda(k)-\varepsilon_\lambda(k')]= \\ \left[\int_0^{k_0}dk'+\int_{k_0}^\infty dk'\right]d\phi k'g(k')\delta[\varepsilon_\lambda(k)-\varepsilon_\lambda(k')]\\=
\left[\int_0^{k_0}\frac{d\varepsilon'}{\left|\frac{\partial{\varepsilon'}}{{\partial k'}}\right|}+\int_{k_0}^\infty \frac{d\varepsilon'}{\left|\frac{\partial{\varepsilon'}}{{\partial k'}}\right|}\right]d\phi k'g(k')\delta[\varepsilon'_\lambda-\varepsilon_\lambda],
\end{split}		
\end{equation}
Eq.~\ref{Boltz} becomes
\begin{equation}\label{Boltz1}
\begin{split}
    -e\bm{v}_k\bm{E}\frac{\partial f}{\partial\varepsilon}=\sum_{\lambda'=\pm}\sum_{i=1}^2\frac{D_i(\varepsilon)}{2\pi\hbar^2}\int d\phi_i|\langle\lambda_i, \bm{k}_i|V|\lambda, \bm{k}\rangle |^2\\ \times [f(\lambda_i,\bm{k}_i)-f(\lambda,\bm{k})],
\end{split}		
\end{equation}
where $\bm{v}_k=(1/\hbar)(\partial{\varepsilon_{k}}/{\partial\bm{k}})$  and the DOS for the $k_i$ branch of the spectrum $$D_i(\varepsilon)=\frac{k_i(\varepsilon)}{|v_{k_i}(\varepsilon)|}=\frac{k_i(\varepsilon)}{\left|\frac{\partial{\varepsilon_{k}}}{{\partial k}}\right|_{k=k_i}}.$$ Here we present the results for the case of a fully occupied valence band, when only scattering in the conduction band is considered, so below we omit the band index $\lambda$.

Linearizing the Boltzman equation in the electric field $\bm{E}$, $f(\bm{k})=f_0(\varepsilon)+ f_1(\bm{k})$, and substituting $$f_1(\bm{k})= -e\bm{E}\bm{\Psi}\frac{\partial f_0}{\partial\varepsilon},$$ we get
\begin{equation}\label{Bolz1}
\begin{split}
    \bm{v}_k=\sum_{i=1}^2\frac{D_i(\varepsilon)}{2\pi\hbar^2}\int d\phi_i|\langle \bm{k}_i|V| \bm{k}\rangle |^2[\bm{\Psi}(\bm{k}_i)-\bm{\Psi}(\bm{k})].
\end{split}	
\end{equation}
For the axial symmetric system the vector $\bm{\Psi}$  can be presented in polar coordinate as $\bm{\Psi}=\chi(k)\bm{k}/k$. Substituting it in Eq.\ref{Bolz1} and multiplying the equation by $\bm{k}/k$ we get
\begin{equation}\label{Boltz2}
\begin{split}
    v_k=\sum_{i=1}\frac{D_i(\varepsilon)}{2\pi\hbar^2}^2\int d\phi_i|\langle \bm{k}_i|V| \bm{k}\rangle |^2\left[\chi(k_i)\frac{\bm{k}\cdot\bm{k}_i}{kk_i}-\chi(k)\right].
\end{split}	
\end{equation}
We substitute into Eq.\ref{Boltz2} for $k$ the values $k_1(\varepsilon)$ and $k_2(\varepsilon)$ of the two branches of the dispersion consequentially and get for each value of energy $\varepsilon$ a set of two equation for $\chi_1(\varepsilon)=\chi(k_1(\varepsilon))$ and $\chi_2(\varepsilon)=\chi(k_2(\varepsilon))$:
\begin{equation}\label{Boltz3}
    v_{k_j}=\sum_{i=1}\frac{D_i(\varepsilon)}{2\pi\hbar^2}^2\int d\phi_i|\langle \bm{k}_i|V| \bm{k}_j\rangle |^2\left[\chi(k_i)\frac{\bm{k}_j\cdot\bm{k}_i}{k_jk_i}-\chi(k_j)\right].
\end{equation}
These equations can be written as
\[
\begin{cases}\label{B1}
    \color[rgb]{0,0,0}Q_{11}\chi_1(\varepsilon)-Q_{12}\chi_2(\varepsilon) =v_{k_1}\\
		Q_{21}\chi_1(\varepsilon)-Q_{22}\chi_2(\varepsilon)=v_{k_2}
\end{cases}
\]		
Here
\begin{equation}\label{Boltz4}
\begin{split}
    Q_{11}=\frac{D_1(\varepsilon)}{2\pi\hbar^2}\int d\phi_1|\langle \bm{k'}_1|V| \bm{k}_1\rangle |^2[\cos(\phi_1-\phi_1')-1]\\
		-\frac{D_2(\varepsilon)}{2\pi\hbar^2}\int d\phi_2|\langle \bm{k}_2|V| \bm{k}_1\rangle |^2;\\
		Q_{21}=\frac{D_2(\varepsilon)}{2\pi\hbar^2}\int d\phi_2|\langle \bm{k}_2|V| \bm{k}_1\rangle |^2\cos(\phi_1-\phi_2);\\
		Q_{22}=\frac{D_2(\varepsilon)}{2\pi\hbar^2}\int d\phi_2|\langle \bm{k'}_2|V| \bm{k}_2\rangle |^2[\cos(\phi_2-\phi_2')-1]\\
		-\frac{D_1(\varepsilon)}{2\pi\hbar^2}\int d\phi_1|\langle \bm{k}_1|V| \bm{k}_2\rangle |^2;\\
		Q_{12}=\frac{D_1(\varepsilon)}{2\pi\hbar^2}\int d\phi_1|\langle \bm{k}_2|V| \bm{k}_1\rangle |^2\cos(\phi_2-\phi_1).
\end{split}
\end{equation}

Thus we got two different equations for $\chi_1(\varepsilon)$ and $\chi_2(\varepsilon)$. Consequently, there are two different nonequilibrium distribution functions for the two MHD branches. The first term in $Q_{11}$ (Eq.~\ref{Boltz4}) is similar to the relaxation time approximation for the case of single-branch dispersion. However there are additional terms: a second term accounting for the rate of scattering from the first branch to the second, and a term $Q_{12}$ describing the scattering from the second branch to the first. The second equation of the system is similar, only the first and second branches are interchanged. Note that to demonstrate the angular dependence of electron scattering in Fig.~3 the integrand of this $Q$-matrix are shown.

We numerically solved the system of equations (Eq.~\ref{Boltz4}) for a fixed energy $\varepsilon$ to determine the functions  $\chi_1(\varepsilon)$ and $\chi_2(\varepsilon)$. While the equilibrium distribution function is identical for both MHD branches, under applied field the occupation of the two branches changes differently. Notably, the group velocity exhibits a key asymmetry: on the inner ring, it is antiparallel to $\bm{k}$, whereas on the outer ring, it is parallel to $\bm{k}$. Consequently, the corrections to the distribution function—governed by $\chi_2(\varepsilon)$ near the inner edge and $\chi_1(\varepsilon)$ near the outer edge—acquire opposite signs.
 
Having $\chi_1(\varepsilon)$ and $\chi_2(\varepsilon)$ we can define the current in the system. We suppose that electric field is directed along $OX$ axis. Then the amendment to distribution function is a vector consisting of two components: $$f_{11}(\varepsilon)= -eE\chi_1(\varepsilon)\frac{\partial f_0}{\partial\varepsilon}\cos\phi$$ nearby the inner ring and $$f_{12}(\varepsilon)= -eE\chi_2(\varepsilon)\frac{\partial f_0}{\partial\varepsilon}\cos\phi$$ nearby the outer ring. The velocity along $OX$ axis for i-th branch  of MHD equals $v_{k_i}\cos\phi_k$. Then the current  equals:
\begin{equation}
\begin{split}
J=-2e\int\frac{d^2k}{(2\pi)^2} \cos\phi \left[v_{k_1}f_{11}(\varepsilon)+v_{k_2 }f_{12}(\varepsilon)\right] = \\ \frac{e^2E}{\hbar^2}\int\frac{ d\varepsilon}{2\pi} \frac{\partial f_0}{\partial\varepsilon}\left [D_1(\varepsilon)\chi_1(\varepsilon)\frac{\partial{\varepsilon_{k}}}{{\partial k}}|_{k=k_1}+D_2(\varepsilon)\chi_2(\varepsilon)\frac{\partial{\varepsilon_{k}}}{{\partial k}}|_{k=k_2}\right]
\end{split}
\end{equation}
where 2 is for two spins.
\begin{figure}
    \centerline{\includegraphics[width=1.\linewidth]{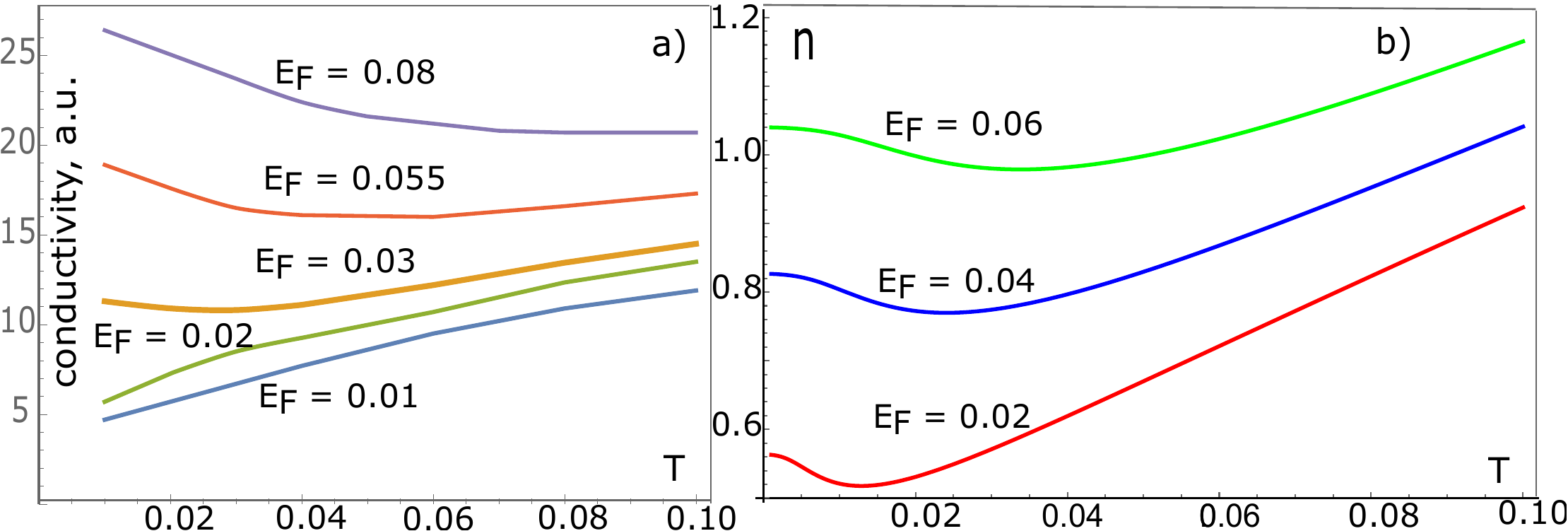}}
    \caption{Dependence of current (a) and carrier density (b) on the temperature for given Fermi levels shown near the curves.}
    \label{fig4}
\end{figure}

Figure 4a shows the temperature dependence of the current for a fixed Fermi level. It is seen that at the low Fermi energy the current increases with the temperature. For such a low Fermi energy and low temperature the carrier density dependence on temperature is non-monotonic due to the strong change of DOS with energy near singularity (see Fig.4b ). At some temperature close to the fixed Fermi energy, the carrier density begins to increase with temperature rapidly. The screening of the small angle scattering potential also depends strongly on the temperature for such Fermi levels, namely the scattering potential increases with temperature (see Fig.2a). But the increase of carrier density is faster and the current increases with the temperature. When the Fermi level is higher the dependence of the density on the temperature is weaker, the current decreases with the temperature, the main reason of a current decrease being the increase of the impurity potential due to the decrease of screening with the temperature.

\begin{figure} \label{fig5}
    \centerline{\includegraphics[width=1.\linewidth]{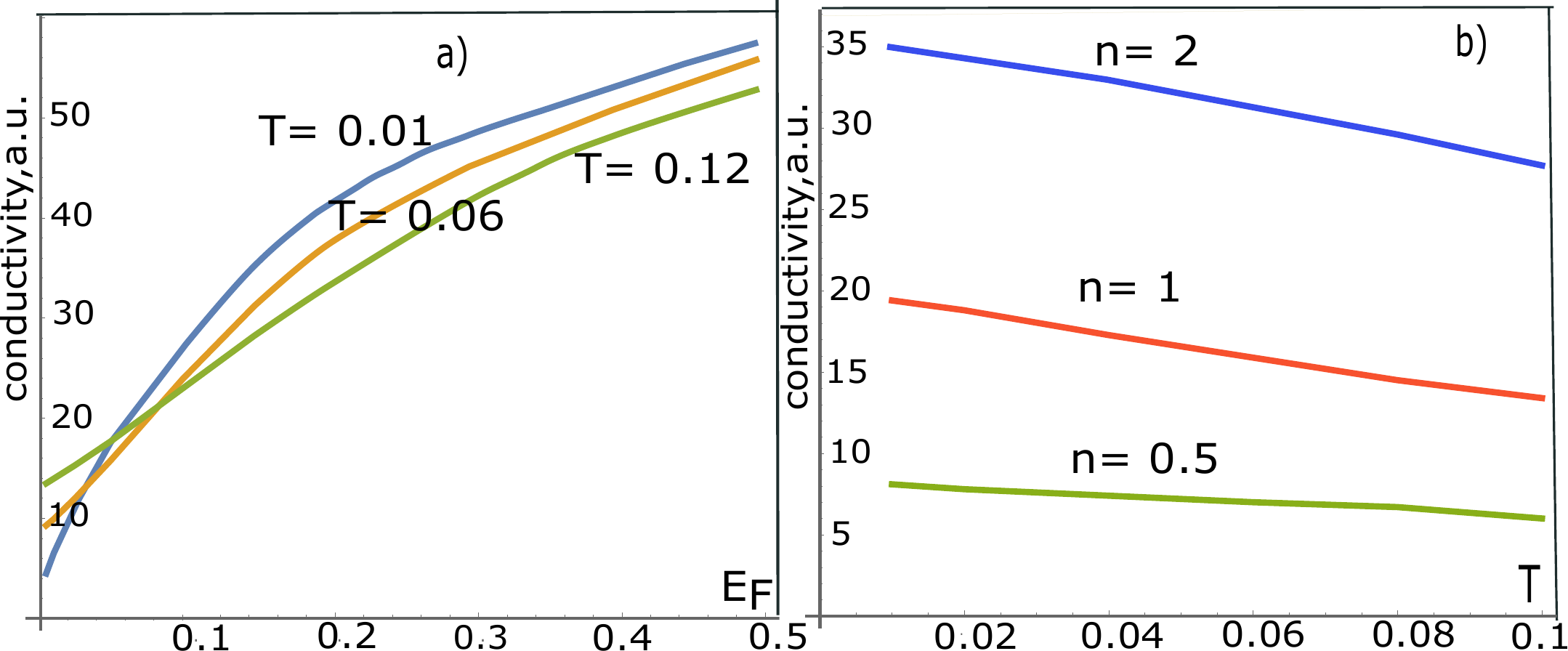}}
    \caption{Dependencies of the conductivity on the Fermi level for fixed temperatures (a) and temperature dependencies of the conductivity for fixed electron densities (b).}
\end{figure}

The dependence of the current on the Fermi level for fixed temperatures are shown in Fig.5a. Conductivity increases with increasing Fermi level because both the electron density and the screening of the impurity potential increase with increasing Fermi level. Fig.5b shows the temperature dependencies of the conductivity for fixed electron densities. It is seen that the conductivity decreases with the temperature. This is a result of a weaker screening of the charged impurity potentials at higher temperature for a fixed electron density.

\section{Conclusion}\label{S_conclusion}

In conclusion, we have studied how the essential properties of the band states in MHD materials, namely the ring-shape Fermi surface, quantum metric of the basis band states, and the  electron–electron interaction affect the temperature and electron density dependencies of the conductivity. The main effects are originated from the changes in the scattering potential: the presence of two Fermi contours gives rise to three modes of Friedel oscillations, the effect of which increases when the Fermi energy is close to the singularity points of the DOS. Changes in the scattering potential allow us to qualitatively explain the non-trivial features of the dependencies of conductivity on the temperature and electron density.

A remarkable consequence of the presence of ring-shaped Fermi surfaces is that the momentum in such systems is not a single-valued function of energy and therefore the electron transport cannot be described within the usual approximation of the momentum relaxation time. We have proposed a correct approach to solving the Boltzmann equation for the distribution function in this case and studied in detail the consequences of this feature for two-branch dispersion. This approach can be extended to more complex problems, including extrinsic skew scattering and other transport phenomena.

\begin{acknowledgments}
This work was carried out in the framework of the state task for the Kotelnikov Institute of Radio Engineering and Electronics.
\end{acknowledgments}
        
\bibliography{scattering}

\end{document}